# Performance Advancement of Wireless Sensor Networks using Low Power Techniques and Efficient Placement of Nodes

**Fatma Almajadub and Khaled Elleithy**
*Computer Science and Engineering Department*, University of Bridgeport, CT-06604, USA
falmajad@bridgeport.edu, elleithy@bridgeport.com

*Abstract:-*
In this paper, we present optimization techniques for WSNs. Our main goal is to minimize the power consumption and latency. We address the problem of minimizing the energy consumption in WSNs including hardware. ZigBee protocol is used to design nodes on WSN to achieve a very low power consumption rate. Furthermore, we propose to use IRS protocol in WSN within a ZigBee technique to discover information from unaware locations and achieve efficiency of energy and sacrifices latency. Our main idea is to support WSNs with both ZigBee technique and IRS protocol. In addition, we address the problem of efficient node placement for congestion control in WSNs. Thus, we evaluate the performance of specific routing and some algorithms of congestion control when wireless sensor nodes are deployed under different placements of network. To demonstrate the strength of the used algorithms, our simulation in C# proves that ZigBee-IRS- ESRT-Flooding approaches reduce the power consumption from 10% to 19% when compared to existing techniques of low Power and node placement.

## I. INTRODUCTION

In today's world, the development of Wireless Area Networks (WANs) makes wireless communications a current trend. However, the development of sensor techniques that are highly effective in transmitting and receiving data makes Wireless Sensor Networks (WSNs) a plausible platform of communications that is cheap and easy to deploy. Therefore, optimization of power consumption helps boost the performance of WSNs. The most challenging factor in optimizing WSNs is to achieve low energy consumption and low latency for reliable data communications with efficient node placement. In this paper, we provide a framework to enhance the performance of WSNs. The major designs and algorithms used in this framework are discussed in the following section.

ZigBee is a design that specifies the maximum data rate of its link by around 250 kbps, which is not enough with the increasing requirements of data transmission rate. [1]-[2] This design can be used to develop ubiquitous applications. It has the following advantages:
1) ZigBee is a wireless mesh network which provides a larger range and high reliability. Therefore, it has great advantages over high fault tolerance, flexibility, and autonomy. [3]
2) ZigBee is the simple technology that was designed with a cost less than other wireless personal networks which makes it more usable in the monitoring and controlling applications. For example, ZigBee is used in the radio bands of scientific and medical applications. [1] In addition, the specifications of ZigBee are freely available for all non-commercial purposes. [1]-[3]
3) ZigBee is designed with low power which makes smaller batteries last longer. ZigBee's other quality is the fact that it can wake up and turn to active mode in only 15ms or less. In addition, not only can its latency be very low but also the devices with ZigBee can be very responsive particularly when it is compared to other technologies like Bluetooth which have delays of 3 second to turn of sleep mode to active mode. However, the average power consumption of ZigBee is very low because ZigBee was designed with abilities that enable it to sleep most of the time. This quality makes the battery last longer.

Increasing Ray Search (IRS) Protocol was proposed in [4]. This Protocol considered the problem of information discovery in WSN. IRS is designed to achieve an energy efficient and scalable search. The goals of this protocol are as follows:
1) Achieve efficiency of energy and sacrifices latency.
2) Route the packet along a set of tracks which are called rays to organize them. When the packets are sent along all these rays, the whole area will be allocated for it. This protocol with these rays minimizes the overlap of the transmissions.
3) These rays maximize the probability of discovering the target information by consuming the least amount of energy.

There are many algorithms that help to achieve good nodes placement; two of them are the algorithm of reliable data transport (ESRT) and the algorithm of a generic routing (Flooding). Both algorithms have been presented in [5], [6] [7] and [8] only for the comparison between them. Here we use them to improve the performance of WSN so that these algorithms use the multiple paths as alternative paths in order to forward the surplus traffic from source to sink by using specified placements.

The rest of this paper is organized as follows: section II presents the proposed work; section III discusses the implementation while section IV provides the simulation results. In the final section, we provide the conclusion and future works.

## II. PROPOSED FRAMEWORK

### A. Low Power Optimization

In WSNs, most studies and researches aim at minimizing power consumption in the entire network; however, these studies and researches are subject to some constraints of quality of service (QoS). Therefore, in this part of the paper which deals with low power optimization, we aim to address and handle the problem of minimizing the full energy consumption in WSNs, which handles the energy consumption of the hardware. Hence, we need to know the following:
1) Basic properties of the optimal power distribution.
2) General problem of energy minimization which is based on System parameters like the number of antennas (N).
   a) The sending strategy which includes the number of parallel streams of data and beam formers.
   b) The sending powers.
3) We need to consider the relation between sending time and sending power in order to optimize both jointly as well as find a good trade-off between optimal energy and optimal power-time.
4) The analytical model of energy consumption in WSNs where nodes are constrained power.

#### i. Required Performance measurements

1. Probability of finding the target information:
   We need to measure the probability of finding the target information in order to measure the successful probability and non-determinism of the search protocols.
2. Consumed Energy:
   We need to measure the total consumed energy in the network in order to find the target information.
3. Latency:
   We need to measure the time taken in order to find the target information. We can calculate the time difference between the time at which the sink node starts the search by sending the packet, and the time at which the target node receives the packet.
4. The number of sent and received bytes:
   We need to measure the average number of sent and received bytes in the network in order to find the target information.
5. The number of sending bytes:

We need to measure the average number of sending bytes in a WSN network in order to find the target information. According to that we measure this number instead of measuring the number of sending messages which doesn't have a uniform format within the protocols.

#### ii. Mechanism of Reducing Power Consumption within WSNs:

According to [4]; a mechanism for reducing power consumption in WSNs is as follows:

1) An optimized path is prepared for a sink node which is set by the use of a common channel in which the first and second nodes use a CSMA scheme.

2) A first channel is set in which sending and receiving slots are specified and allocated for sending and receiving packets.

3) During the first sending slot a packet is sent to the second node by using a TDMA scheme.

4) Within the first set amount of time, if a packet is not received from the second node during the first received slot, the first received slot is allowed to go through transition and send an inactive state.

5) The first node is one of the sink nodes. It is at least one of the parent nodes as well as one of the child nodes of that parent node while the second node is one of the child nodes of the first node.

#### iii. The Proposed Protocol of an Energy-Efficient Routing in WSNs

According to [4] the Low- ad hoc network doesn't address the features of WSNs correctly, thus on WSNs, it is not enough to reduce and minimize full energy consumption. We need to maximize the lifetime of the whole WSN network, after which the entire network connectivity can be maintained as long as possible. We proposed considering various scenarios of multi-hop to achieve following:

1) Calculate the energy of each bit.
2) Calculate the efficiency and consumed energy of each node, and afterwards, for the whole network.

Here, we need to analyze the consumed energy at each node by the radio. These analyses should use all detailed models. In addition, we must study, know and understand the Multi-hop topologies with optimal spacing of nodes. Therefore, the numerical computations will show the effects of packet routing as well as the effects of medium access control and coding. We should know that the message usage of a simple multi-hop system is based on a strategy that is not always the best.

Furthermore, we proposed a new design of WSN which is called ZigBee_IRS. We use ZigBee to design the nodes of WSN and we try to improve it in the simulation by increasing the maximum data rate of its link to around 1000 kbps which should be enough with the increase of the data transmission rate within the WSN network. Using 1000kbps will be better than 250 Kbps, which is used now in the real world. To be more specific, we planned to combine the IRS protocol which was proposed in [4] with the design of ZigBee to create the new design of ZigBee_IRS and then add it to the proposed WSN network in order to determine optimal energy number of parallel streams of data per link for a certain SIR requirement and make it work within it. Our main idea of ZigBee_IRS is "when protocol IRS makes a subset of total sensor nodes on WSN it sends a packet to cover the full area and makes other nodes listen in order to achieve low power consumption. Here, ZigBee will make these listening nodes sleep during the period of listening; thus, the ZigBee technique with IRS protocol within WSNs achieve low power consumption and optimize the performance of WSN. We will try proving this by simulation; we think this combination will achieve better results.

## B. Optimizing the Placement of nodes

Sensor nodes in WSNs act group are very small devices that communicate by using a wireless transmitter. These nodes are cooperated autonomously to form a logical WSN network. Data packets in this logical network are routed towards base-stations which are called management nodes or Sinks. These packets are routed by the system of hop-by-hop[5]. Thus, WSNs consist of a large set of these nodes which are distributed within a wide geographical area to perform two specific functions; the sensing and monitoring within this area in order to achieve some services such as traffic and environmental monitoring, efficient industry production, and security at home. However, WSNs with these nodes operate within a light load and have a sudden response to the detected events or monitored events.

WSNs were used in military applications but nowadays, WSNs are used in civilian applications. For example, they are used in health monitoring, habitat observation and object tracking...etc. [5] The performance of WSNs is affected by heavy traffic load. We can say this problem worsens in low powered WSNs and unreliable WSNs.

Hence, we can say that the eminent thing about WSNs is the way these sensor nodes are placed within it on the monitored field of WSNs; it can improve the performance of WSNs or worsen it. Suitable node placement is fundamental to guarantee good sensing and communication.

In this part of the paper, which discusses the optimization of placement nodes, we will try to analyze and compare the performance of only two algorithms of congestion control over efficient placement of nodes.

## i. Congestion control in WSNs

- **The Kind of Congestion in WSNs**

In WSNs where congestion occurs within a temporarily high load, nodes cannot carry this load when it is higher than what they can process in certain time. According to [5], there are two kinds of congestions in WSNs:

1) Congestion of the medium:
This congestion occurs in the medium when two nodes or more send data at the same time, thus a collision of packets occurs. It can be handled by some improvements in the MAC layer.
2) Congestion in the Queue or Buffer:
This congestion occurs in the queue or the buffer. Each node has a queue or a buffer that is used in sending and receiving packets. However, when this buffer holds packets to be sent, the dropping of packets or overflows might occur; particularly, when WSNs are operated with low power. This can be handled by certain improvements in the upper layers (e.g. network layer or transport layer). There are many algorithms that were used as a solution for the congestion problem in WSNs. In the next section we propose to the use of two of such.

- **The Proposed Algorithms for efficient Node Placements in WSNs**

In this paper, we present general ways for the placement of nodes on WSNs. We noted that there are several differences in the performance of the routing algorithms for congestion control over these placements of nodes whether they are traffic control, resource control, reliable data transport or multiple path creation algorithm. Therefore, we propose using the following algorithms:

1) The ESRT "reliable data transport" algorithm. This algorithm is known as an end-to-end algorithm and it focuses on the reliability between sensor nodes of source and sink. This algorithm runs on the sink of WSN when there is network congestion.
2) The Flooding algorithm which is generic routing algorithm. It achieves better results when it is applied to the WSN.

In the simulation, we combined the features of these algorithms in the nodes' settings on the WSN. It was provided with the proposed design of ZigBee_IRS in order to get efficient node placements in WSNs.

These algorithms have been presented on [5], [7] and [8]. However, in this paper, we try to provide the results of simulation of both of these algorithms over different placements. They use multiple paths as alternative paths in order to forward the surplus traffic from source to sink by using specified placements.

Although the Flooding algorithm is not exactly designed for solving the problem of congestion in WSNs, we can use it to reduce high load on the network. This algorithm achieves the purpose of two techniques: traffic and resource controls. Thus, we can also use it to provide multiple disjointed paths. Furthermore, we expect the results of simulation to achieve efficient placement nodes within these algorithms.

### III. IMPLEMENTATION

We have developed a simulation of the proposed work in c# language by using Visual Studio 2012. The simulation allows us to judge the real time behaviors of the network nodes with the given parameters as a basis for evaluating performance enhancements by simulation.

Choosing parameters like stack size, data range and number of nodes allows us to define the node network. Below is a screen shot that shows the selection of various parameters required to set up a node network. Using these parameters, we can define the node architecture, the signal strengths and the behavior of the node network. This enables us to generate the node architecture in the simulation. Figure 1 shows the interface of Node network setting.

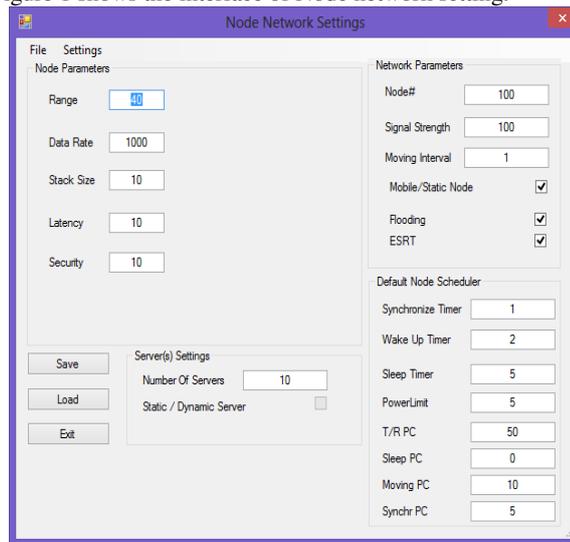

Figure 1: Show The Node Network Setting.

The screen shots below show the generated node structure. Live, when the simulation is run.

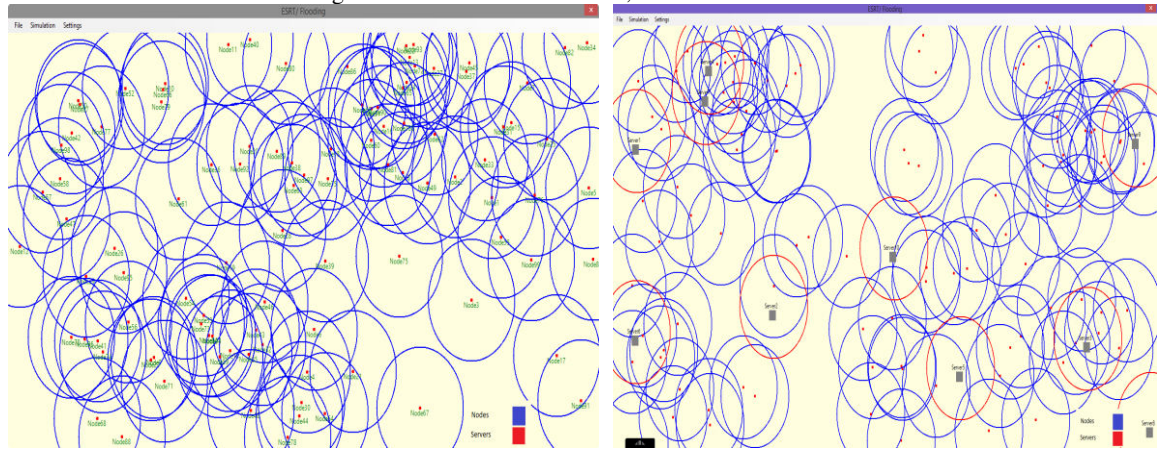

Figure 2(a): Shows The generated live nodes structure of running the simulation   Figure 2(b): Shows the Configuration of WSN with the node network setting.
.

## A. The Implementation of ZigBee_IRS design for Optimization of Low Power: [4]

In this section we discuss the Implementations and results of the ZigBee_IRS design in WSN network within our proposed work.

The configuration of WSN that is implemented in Figure 2(a) and (b) differ from other WSNs as mentioned in section II. On WSNs, the sensor nodes are provided with specific requirements like low memory, limited energy availability, reduced processing power, and energy efficiency which is a main issue for designing the routing for WSNs or any network. Routing protocols of WSNs are based on clustering; their cluster heads are selected randomly to execute many tasks. Thus, this approach sometimes causes lopsided consumption of energy as well as short lifetime of the network. We discovered by simulation that the proposed design of ZigBee_IRS which uses the combination of the ZigBee design and IRS protocol makes a cluster head of each subset WSN and distributes the load of energy equally into its members, depending on their usage of energy by the chain technique where the cycle is different in a cluster. We found that the results of this technique's simulation or design gave the best performance of power management as well as prolonged the lifetime of the network. Therefore, we inferred the following:

**1)** Optimizing the energy consumption in WSNs on our simulation was by using systems of multiple antennas which are able to provide transmissions with a high data rate that we proposed to be 1000 Kbps unless in a fading environment where it has a different data rate. We found that these systems achieved that rate without the need to increase the signal bandwidth.

**2)** By understanding how to minimize the sending power, which is subject to the requirements of the SIR sensor intelligence routing protocol, we can solve the problem of energy minimization which is also subject to the requirements of SIR and is equal to the problem of minimizing transmitted power.

**3)** By adding IRS protocol to a fixed number of antennas and a fixed transmission time as well as formers of the fixed beam at each pair of transmitter and receiver, the minimization problem of energy will be subject to the SIR requirements which is equivalent to minimizing powers of transmission which are subject to the same SIR requirements.

**4)** By considering the relation between transmission time and transmitted power as well as optimizing both jointly, we found that there is a trade-off between the time of an optimal power and low energy. Therefore, we proposed a design of ZigBee_IRS that determines the energy-optimal number of data streams per link for a certain SIR requirement.

**5)** By adding the design of ZigBee, we can solve most of the Power issues in WSNs so that the radio consumes a vast majority of the network energy. Thus, this consumption can be reduced by decreasing the output power of the transmission or by decreasing the duty cycle of the radio. In fact, both of these alternative ways are involved with sacrificing metrics of other systems.

Figure 3(a) compares the average power consumption within the ZigBee_IRS design with the power consumption of the proposed design.

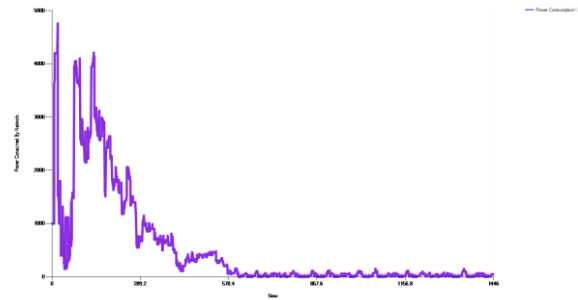

Figure 3(a): Show Average Power Consumption within ZigBee_IRS Design of WSN.

Figure 3(b) shows the latency of the ZigBee_IRS design by comparing the transferred packets (throughput) and lifetime with the proposed design.

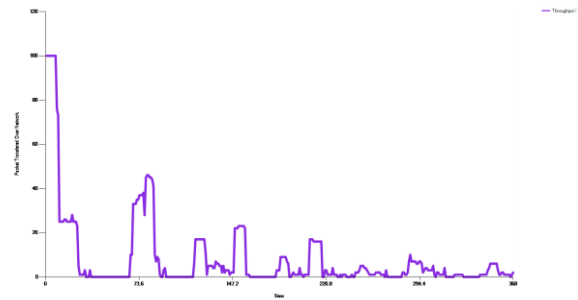

Figure 3(b): Show latency within ZigBee_IRS Design of WSN.

## B. The Implementation of ESRT and Flooding Algorithms for Optimizing the Placement of nodes:

In this section we discuss the Implementations and results of using ESRT and Flooding algorithms for efficient placement of the nodes in WSN network within our proposed work.

By checking the option of ESRT in Figure 1 without checking the Flooding option, we configure the WSN with ESRT only. The implementation of this algorithm is achieved by the following:
1) Regulate report frequency or data rate of all sensors by using the same value.
2) Relieve the congestion on WSN so that the average of data rate is minimally affected by node placements, and the same happens with packet drops.
3) The Packet's delivery delay is a parameter whose behaviour should be related with nodes placement.
4) By applying the ESRT algorithm, efficient nodes placement can be achieved as well as the reduction of the mean time for the transmission of packets.
5) The ESRT presents the same behaviour as SenTCP algorithms [6]. It presents better performance when it is run on Simple Diffusion and Biased- Random placements. The reason is as follows:

a) These are some placements which provide a bigger number of paths from source to sink.
b) When the nodes that form the initial paths are totally power exhausted, the network will still be able to find other paths to forward the data to sink.

Figures 4(a) and (b) show the result of ESRT algorithm on the simulation.

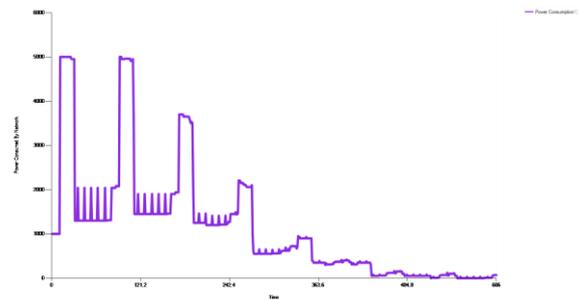

Figure 4(a): Shows average Power consumption of ESRT Algorithm within ZigBee_IRS design on WSN.

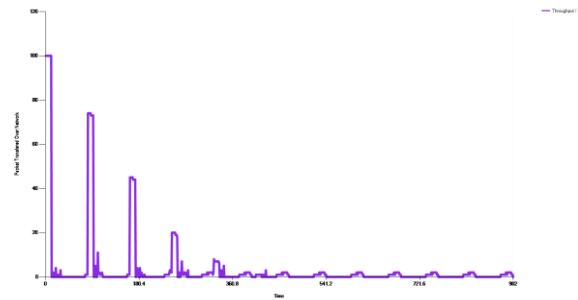

Figure 4(b): Shows latency of ESRT Algorithm within ZigBee_IRS design of WSN.

By checking the option of Flooding in Figure 1 without checking the ESRT option, we configure the WSN with Flooding algorithm only. The implementation of this algorithm is achieved by the following:

1) Each node forwards each message to every node that is in its radio range. Placing the nodes with fewer paths limits the number of packets in the network and therefore fewer drops appear.
2) The source's data rate remains the same because in case of congestion, the flooding algorithm does not implement any "traffic control" functionality.
3) Dropping of packets occur and then there is an increase while time evolves.
4) The flooding will fill the whole network with multiple copies of the same packet.
5) Each node transmits each packet to all of its children. This leads to an amount of transmission bigger than the nodes that have many children around the source. Therefore, these nodes get power exhausted. This will lead to the creation of a "hole" around the source and network "stalls".
Figures 5 (a) and (b) show the result of ESRT algorithm on the simulation.

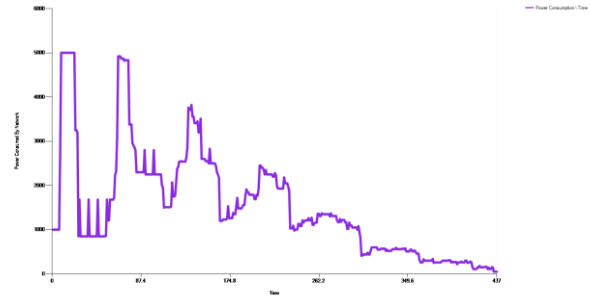

Figure 5(a): Shows average Power consumption of Flooding Algorithm within ZigBee_IRS design of WSN.

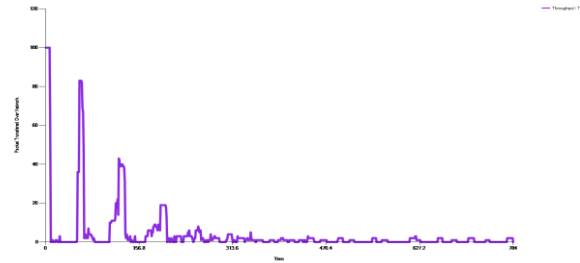

Figure 5(b): Shows latency of Flooding Algorithm within ZigBee_IRS design of WSN.

On the other hand, by checking the option of Flooding and ESRT at the same time in Figure 1, we will configure the WSN with ESRT and Flooding algorithm together. The implementation of combining these algorithms achieves the best performance on the network, as it is shown in Figure 6(a) and (b).

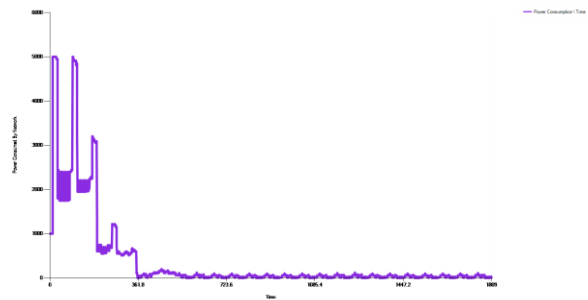

Figure 6(a): Shows average power consumption within ESRT and Flooding Algorithms combined together on WSN.

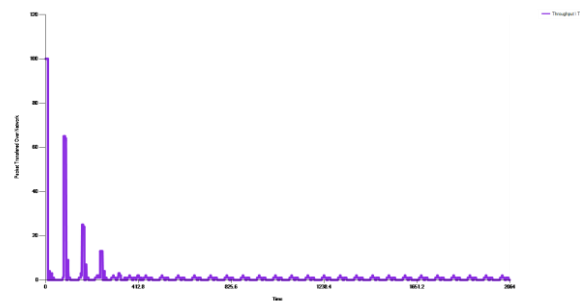

Figure 6(b): Shows latency of ESRT and Flooding Algorithms combined together on WSN.

As a result, by using these algorithms and comparing the placements of nodes, we inferred the following:

**a)** The placements which present the worst performance should present the best performance when the flooding and ESRT are applied together.

**b)** The placements used have fewer nodes around the source.

However, it is the same thing in the case of the packet's delivery delay; the placements that create fewer paths are able to forward the data sooner. Furthermore, it requires checking the remaining percentage of the network's energy when the network stops.

## IV. THE SIMULATION RESULTS

In this paper we conclude that we can successfully develop a novel algorithm to improve the power consumption and the latency of the WSN network. The figure below shows the power consumption of a scenario over the WSN network in real time.

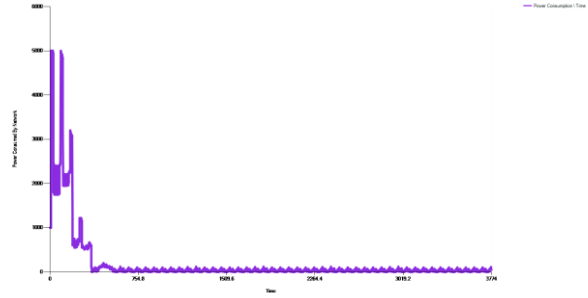
Figure 7(a): Shows average Power consumption within design of WSN.

We can also measure the latency in our network in real time. The following graph shows the latency in the network:

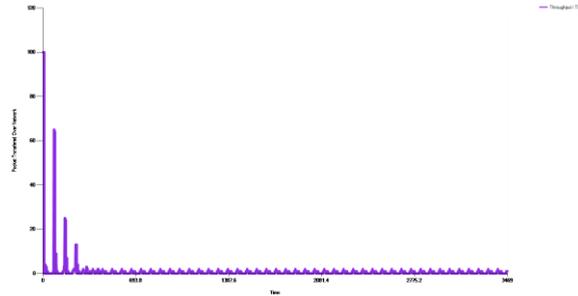
Figure 7(b): Latency within design of WSN.

Thus with the generated results we can say that we have successfully proven that our proposed work allows improvement of the power consumption and the latency of a wireless sensor node network.

## V. CONCLUSIONS

In this Paper, we have presented the optimizations of performance in advancing WSNs over two important sides. On one side, we addressed the problem of minimizing the energy consumption in WSNs, including energy consumption for hardware. This was done by adding the ZigBee technique to design nodes on WSN so that it will make the nodes of WSN able to wake up and turn to active mode in only 15ms or less. While ZigBee is designed for low power, it achieves the very low power consumption because it is designed with abilities that make the nodes able to sleep most of the time, which results in the battery lasting longer. According to this quality, we proposed providing IRS protocol to the WSN that was designed with the ZigBee technique, so that this protocol would be used for discovering information of unaware locations and achieving efficiency of energy and a sacrifice of latency. In addition, when protocol IRS makes a subset of total sensor nodes on WSN, it sends a packet to cover the full area, and it makes other nodes listen in order to achieve low power consumption. Here ZigBee makes these listening nodes sleep during the period of listening, thus the ZigBee technique with IRS protocol within WSNs achieves low power consumption and optimizes the performance of WSN. In the simulation, we used the data rate of 1000Kbps rather than 250 Kbps which is used in real world, thus we inferred that 1000Kbps is enough for the increase of data transmission rate within the network. In addition, in the simulation we considered the relations between sending time and sending power, and we optimized both jointly to achieve a good tradeoff between energy and optimal power-time. On the other side, we addressed the problem of efficient node placement for congestion control in WSNs. We also evaluated the performance of specific routing and some algorithms of congestion control when wireless sensor nodes are deployed under different placements of network. The algorithms we examined are ESRT [6] and Flooding [5]. ESRT algorithm represents a certain category of congestion control with the placement of nodes, thus it is a routing algorithm. On the other hand, Flooding algorithm worked with the fact that each node always tries to send every message to every neighbour node. In the simulation, we inferred that both algorithms employ multiple and alternative paths, they use data in sending and receiving from source to sink so that this transmission is significantly favored by denser placements of nodes, either at the source or the sink of WSN, since these nodes within these algorithms can create many paths. Therefore, fewer packets are dropped while these algorithms try to extend the lifetime of the whole network. However, these algorithms employ the shortest path for sending and receiving packets from source to sink in order not to be affected by the placements of different nodes at continuous heavy load in the network, but these algorithms help to present the shortest lifetime of a network. In this paper, we have presented the power issues of WSN; thus we presented the role of ZigBee and IRS protocol to reduce the power consumption on WSN. Finally, for future work we can say that the vast majority of the system's energy in a wireless sensor node is consumed by the radio. We can reduce this consumption through either decreasing the transmission output power or decreasing the radio duty cycle. Each of these alternative ways includes some manipulating and sacrificing some other system metrics.

### ACKNOWLEDGEMENTS

The authors of the paper would like to acknowledge Mrs. Azza Al Shaer for her valuable comments in editing this paper.

The authors of the paper would like to acknowledge Mrs. Doreen and Mrs Lydia Iarocci for review the language of this paper.

## REFERENCES

[1]. Dr. Debmalya Bhattacharya, R.Krishnamoorthy, "Power Optimization in Wireless Sensor Networks", IJCSI , Vol. 8, Issue 5, No 2, September 2011.

[2]. Liu Yanfei, Wang Cheng, Qiao Xiaojun, Zhang Yunhe, Yu chengbo,Liu Yanfei, "An Improved Design of ZigBee Wireless Sensor Network", IEEE 2009. R. E. Sorace, V. S. Reinhardt, and S. A. Vaughn, "High-speed digital-to-RF converter," U.S. Patent 5 668 842, Sept. 16, 1997.

[3]. Chia-Ping Huang, "Zigbee Wireless Network Application Research Case Study within Taiwan University Campus", Proceedings of the Eighth International Conference on Machine Learning.

[4]. Surendra bilouhan, Prof.Roopam Gupta, "Optimization of Power Consumption in Wireless Sensor Networks", IJSER © 2011 http://www.ijser.org.

[5]. Charalambos Sergiou and Vasos Vassiliou, University of Cyprus, University Ave, 2019, Nicosia, Cyprus, licensee InTech. This is an open access chapter distributed under the terms of the Creative Commons Attribution License (http://creativecommons.org/licenses/by/3.0), INTECH ©2012.

[6]. Y. Sankarasubramaniam, O. Akan, and I. Akyildiz. ESRT: Event-to-Sink Reliable Transport in Wireless Sensor Networks. In MobiHoc '03: Proceedings of the 4th ACM International Symposium on Mobile Ad Hoc Networking & Computing, pages 177–188, New York, NY, USA, 2003. ACM.

[7]. V. Vassiliou and C. Sergiou. Performance Study of Node Placement for Congestion Control in Wireless Sensor Networks. In New Technologies, Mobility and Security (NTMS), 2009 3rd International Conference on, pages 1 –8, Dec. 2009.

[8]. Charalambos Sergiou and Vasos Vassiliou. Energy utilization of HTAP under specific node placements in Wireless Sensor Networks. In Wireless Conference (EW), 2010 European, pages 482 –487, 12-15 2010.

[9]. C. Wang, K. Sohraby, and B. Li. SenTCP: A Hop-by-Hop Congestion Control Protocol for Wireless Sensor Networks. IEEE INFOCOM(Poster Paper), March 2005..

[10]. Charalambos Sergiou, Vasos Vassiliou, and Andreas Pitsillides. Reliable Data Transmission in Event-Based Sensor Networks During Overload Situation. In WICON '07: Proceedings of the 3rd International Conference on Wireless Internet, pages 1–8, Austin, Texas, October 2007.

[11]. Mohamed Younis and Kemal Akkaya. Strategies and Techniques for Node Placement in Wireless Sensor Networks: A Survey. Ad Hoc Networks, 6(4):621–655, 2008

[12]. Sameer Tilak, Nael B. Abu-Ghazaleh, and Wendi Heinzelman. Infrastructure tradeoffs for sensor networks. In WSNA '02: Proceedings of the 1st ACM international workshop on Wireless sensor networks and applications, pages 49–58, New York, NY, USA, 2002. ACM.

[13]. Mika Ishizuka and Masaki Aida. Performance Study of Node Placement in Sensor Networks. ICDCSW '04: Proceedings of the 24th International Conference on Dis`tributed Computing Systems Workshopsd, pages 598–603, March 2004.

[14]. Chalermek Intanagonwiwat, Ramesh Govindan, Deborah Estrin, John Heidemann, and Fabio Silva. Directed Diffusion forWireless Sensor Networking. IEEE/ACM Transactions on Networking, 11(1):2–16, 2003.

[15]. Ayad Salhieh Jennifer Weinmann _ Manish Kochhal _ Loren Schwiebert, "Power Efficient Topologies forWireless Sensor Networks" reference.kfupm.edu.sa/.../power_efficient_topologies_for_wireless_529104.pdf.

[16]. Shaosheng Li, Wenze Li, Jianxun Zhu, "A Novel Zigbee Based High Speed Ad Hoc Communication Network", Proceedings of IC-NIDC2009,IEEE 2009.

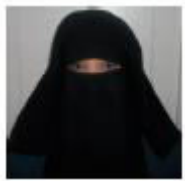

**Mrs. Fatma Almajadub** is a Master student in the School of Engineering and Computer Science at the University of Bridgeport. She has finished her Bachelor's degree at Sabha College for Computer Science in 2004-2005. She graduated with highest honors, ranking 4th in her Bachelor degree and awarded merit certificate. Mrs. Almajadub worked as a teacher at Sabha University in 2004-2009. She worked also as teacher in Algamaheria institution to high education since 2006-2009. Mrs. Almajadub is interested in programming, network area, mobile communication, and some software applications.

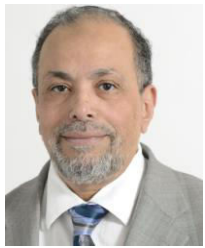

**Dr. Khaled M Elleithy** is the Associate Dean for Graduate Studies in the School of Engineering at the University of Bridgeport. He has research interests are in the areas of network security, mobile communications, and formal approaches for design and verification. He has published more than two hundreds fifty research papers in international journals and conferences in his areas of expertise. Dr. Elleithy is the co-chair of the International Joint Conferences on Computer, Information, and Systems Sciences, and Engineering (CISSE). CISSE is the first Engineering/Computing and Systems Research E-Conference in the world to be completely conducted online in real-time via the internet and was successfully running for six years. Dr. Elleithy is the editor or co-editor of 12 books published by Springer for advances on Innovations and Advanced Techniques in Systems, Computing Sciences and Software.